# A Taxonomy of Adaptive Traffic Signal Control

Andalib Shams, A. M. Tahsin Emtenan, and Christopher M. Day, *Member, IEEE*

*Abstract*—Research on adaptive traffic signal control (ATSC) extends back to at least the 1960s, and many ATSC methods have been proposed over the years. This paper provides a review of this research and proposes a taxonomy for organizing it, accompanied by a consistent vocabulary for discussing the control concepts. We begin from the well-established concept of control generations. Next, we classify the ATSC methods according to their topographic structure (local-only, system/hierarchical), time resolution of decision-making (continuous versus planning-horizon), type of decision (rule-based or optimization), objective function, cyclic/acyclic nature, and additional subcategories relevant to certain "families" of methods. These various elements of system control are organized into a taxonomy of ATSC to help future researchers understand the wide diversity of algorithmic approaches to the signal control problem that have been proposed to date, and which can be updated or expanded to incorporate future research.

## I. INTRODUCTION

Traffic signal control is an important interdisciplinary problem in transportation engineering. Methods of signal control that adjust signal timing in response to traffic conditions are generally referred to as adaptive traffic signal control (ATSC). Many different methods of ATSC have been proposed, and there are thousands of publications on this subject in the literature. The objective of this paper is to survey the literature on ATSC and develop a taxonomy to organize the literature.

ATSC is a challenging topic to review comprehensively. A search for the phrase "adaptive traffic signal" on Compendex returns 1,767 results when filtered by the controlled vocabulary "traffic signals" (there are over 7,000 results without this filter). Other search phrases are also relevant. It is infeasible to comprehensively examine every publication. However, by examining a broad cross-section of the literature, it is possible to identify certain trends and commonalities.

One difficulty in discussing ATSC in multiple disciplines is the lack of a consistent vocabulary. Even the definition of "adaptive" is not uniform. ATSC could be defined as self-regulating control based on real-time traffic data, but this definition is loose enough to include actuated control, which is not normally considered "adaptive." Alternatively, the term "adaptive" could be reserved only for "real-time" control, similar to the *Traffic Control Systems Handbook* [1], but this excludes adaptive adjustments to conventional timing plans. Some authors do not use the term "adaptive" at all. This paper proposes a vocabulary for discussing ATSC.

Some previous publications have included some relatively extensive reviews of ATSC. Katwijk *et al.* [2] proposed a taxonomy of ATSC, but limited their discussion to local intersection control, and their taxonomy focuses on optimization methods. Papageorgiou *et al.* [3] reviewed control strategies for urban road network, freeway networks and route guidance. Signal control was one subtopic in the broader review, so details of ATSC are not discussed. Stevanovic prepared a synthesis of ATSC deployments [4] which includes descriptions of several methods provided by the vendors. This review focused mainly on commercial ATSC systems.

Some reviews focused on ATSC for a particular situation or a family of strategies. Wei *et al.* [5] surveyed signal control strategies that use Reinforcement Learning (RL) framework. Jing *et al.* [6] and Wang *et al.* [7] reviewed ATSC strategies for connected vehicle environments. A recent review of computational intelligence in traffic signal control was presented by Qadri *et al.* [8], examining papers published between 2015-2020 and classifying them according to several criteria.

The present paper seeks to review ATSC more generally, but to limit its scope to the most useful work for potential field deployment, we focus on methods that can be applied to multiple phases where opposing turning movements are permitted to terminate independently, with the exception of other papers that make significant methodological contributions. The paper begins with a review of basic signal control concepts. Next, we describe the literature search and criteria for paper selection. After this, we present the criteria for the taxonomy. Finally, we present the organization of the works into the taxonomy.

This paper employs numerous acronyms which are explained by Table VII in the Appendix.

## II. BACKGROUND

This section provides a brief description of signal control methods in general and elements of the history of their development.

### 1) Fixed-Time Control

The first traffic signals on public streets were human-operated, with automatic control emerging in the 1920s [9]. The need for coordination of multiple intersections and variation by time of day were recognized from the beginning and are the basis of conventional control methods that are still widely used.

The simplest signals transfer the right-of-way between two streets, under "two-phase" control (Figure 1a). The need to accommodate crossing turns (*e.g.*, the left turns in right-hand drive countries) led to the inclusion of additional phases,

This work was supported in part by Pooled Fund Study TPF-5(483).
A. Shams is with Iowa State University, Ames, IA 50010 USA (e-mail: ashams@iastate.edu).

A. M. T. Emtenan is with University of Maryland, College Park, MD 20742 USA (e-mail: tahsin@umd.edu).
C. M. Day is with Iowa State University, Ames, IA 50010 USA (e-mail: cmday@iastate.edu).

leading to "four-phase" control (Figure 1b). "Eight-phase" control incorporates two rings that permit the opposing crossing turns to start or end independently, subject to phase compatibility constraints (Figure 1c). Eight-phase control is widely used in North America [10]. Elsewhere, stage-based control is more common, which can achieve similar operation although the phase sequence may be more constrained.

In conventional coordinated control, the basic elements that comprise a timing plan are the cycle, offset, and splits (COS). The *cycle* (or *cycle length*) is the amount of time within which all phases must be served. The use of a common cycle length causes these patterns to repeat in successive cycles, as long as the controllers are synchronized. The time difference between each intersection's local timing and a common system reference point is the *offset*. Finally, the share of the cycle length apportioned to each phase is the *split* of that phase.

*2) Actuated Control*

Actuation allows the service of phases and the duration of green to be adjusted in response to changing traffic demands, as measured by detectors. Detectors generally identify when vehicles are waiting for service on a movement, or if there is still traffic being served on a phase currently in service. Because actuation uses a relatively simple set of rules, it is not generally considered to be a form of "adaptive" control.

Under fully-actuated control, all phases are actuated and are served in a constant sequence (with the possibility of skipping phases that have no demand) and with green times varying according to demand. There is no fixed cycle length, but an effective cycle length that is the consequence of phase durations determined by the actuation rules. Some authors have observed that for random vehicle arrivals (such as at isolated intersections), the performance of fully-actuated control is nearly optimal in terms of delay [11].

*3) Actuated-Coordinated Control*

Actuated-coordinated signal control uses the same COS framework as fixed-time control, but the phase durations are adjustable by actuation, subject to constraints that maintain synchronization. As under fully-actuated control, phases may be skipped if there is no demand, and their green times may be truncated if there is more split time than the amount of demand present for the phase in that cycle. This permits other phases to access the yielded time, which can help if their demand exceeds what the split can handle. The coordinated phases receive special treatment. The controller cannot skip the coordinated phases, and it cannot terminate them unless it is within a portion of the cycle where there is sufficient time to serve the split for the other phases and return to the coordinated phases prior to the start of their split. Numerous settings are available to fine-tune this operation. For a more detailed description, readers may consult the *Signal Timing Manual* [10].

In the United States, fully-actuated control without coordination is generally used at isolated intersections or on arterial corridors during low volume conditions. Actuated-coordinated control is very common on arterial streets, while fixed-time control is mostly limited to some central business districts where the need to serve pedestrians during every cycle limits the utility of actuation. Other countries vary considerably in use of actuation.

*4) Advanced Control Methods*

Conventional control methods can provide an efficient operation when the settings are well-timed for expected traffic conditions. Timing plans can be designed to robustly accommodate some amounts of variation [12], but large variations can lead to inefficient operation. Timing plans are usually developed from traffic counts obtained on a particular date, but traffic is highly variable and changes over time, making it difficult to keep timing plans updated.

One important distinction exists within the definition of generations: the use or non-use of COS. Some acyclic control methods do not use COS. However, being acyclic is not a necessary or sufficient condition for being adaptive. Fully-actuated control is acyclic, and some adaptive methods operate within a COS framework.

In the US, a key research effort in the mid-20[th] century was the Urban Traffic Control System (UTSC) project [13], which started in the 1960s and continued until the early 1980s. This project introduced the notion of "generations" of signal control. Table I presents a description of the UTSC control generations. This offers a useful framework for discussing ATSC.

System-level signal timing optimization is a NP-complete problem [14] and is consequently difficult to solve within a fraction of a second. Real-time ATSC strategies must therefore consider a tradeoff between performance and computational efficiency. Detailed knowledge of these strategies is needed to propose further refinements to these strategies or to implement them in the real world. However, the sheer number of ATSC strategies that have been proposed makes this challenging. The UTSC generations are useful for establishing a high-level perspective on ATSC, but they do not offer much classification beyond the relationship of the algorithms to the COS paradigm. To better survey the field of ATSC, this paper presents a taxonomy of adaptive strategies incorporating multiple criteria.

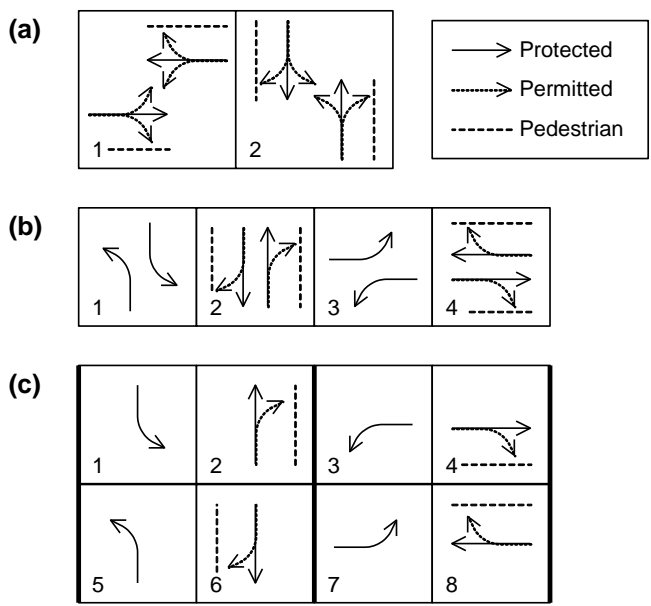

Figure 1 Conventional multiphase signal control: (a) two-phase; (b) four-phase; (c) eight-phase.

TABLE I GENERATIONS OF ADAPTIVE TRAFFIC SIGNAL CONTROL

| Number | Description |
|---|---|
| 0 | **Conventional control.** Groups of signals use COS timing plans that are selected by time-of-day control. Fully-actuated control is used for isolated intersections and sometimes for low volume conditions. |
| 1 | **Traffic Responsive COS Pattern Selection.** Measurements of volume and occupancy are made at various locations throughout the network, and quantities derived from these are used to select from pre-programmed COS timing plans, instead of time-of-day control. |
| 2 | **Adaptive COS Pattern Adjustment.** Similar to conventional control, with the addition of adaptive adjustments that change the timing plans according to measured conditions. |
| 3 | **Real-Time Adaptive Control.** These control methods generally do not use COS, and either replace or augment fully-actuated control with an alternative phase-switching decision method. |

## III. LITERATURE SEARCH METHODOLOGY

To develop a taxonomy of ATSC, we first conducted an extensive literature search. Relevant works were identified by searching for keywords with Google Scholar and Compendex. Additional searches were made with IEEE Xplore, ScienceDirect, arXiv, Scopus, and Web of Science. The following criteria were used to select articles for further review:

- *Clarity*. The underlying signal control algorithm needed to be clearly and adequately explained.
- *More Seminal Work.* Some ATSC methods were discussed by several papers, with some having derivative works. For these families of methods, we selected either the original publication or one that provided the most illuminating description of the control method.
- *Multiphase accommodation*. We focused on methods that are able to handle eight-phase control, with the exception of papers that present novel algorithmic contributions despite being limited to other phasing schemes.
- *Method of evaluation*. Because so much prior work has been demonstrated only in simulation, we did not use consider field evaluation as a criteria for selection. We tended to select works that included comparisons against well-timed and realistic conventional control.
- *Importance of the Work*. The number of citations a paper accumulated was considered a measure of its importance to the ATSC literature. Some papers that were frequently cited by others were included, even if they did not meet all of the above criteria.

The initial search identified over 1700 articles. Using the above criteria, 137 articles were selected for a more detailed read. From that number, 88 ATSC methods were used to develop the taxonomy presented here. The selected articles were then rereviewed to identify characteristics of the control methods, and that information was used to develop a taxonomy.

## IV. MAJOR TAXONOMY ELEMENTS

This section presents findings of the literature search, focusing on criteria that served as the foundation for the taxonomy, which is presented later. More detailed information from the surveyed papers are summarized in Tables IV–VII, which have been included in the appendix.

### A. Control Generation

The UTCS generations are a starting point for classifying ATSC methods. Here we briefly describe developments in advanced signal control relative to each control generation.

#### 1) First-Generation (Traffic Responsive) Control

First-generation control (1-GC) uses measurements of volume and occupancy from detectors distributed throughout the network to select COS patterns from a predefined library. 1-GC control allows these to be selected according to measured demand rather than transitions at arbitrary times of day. The pattern library can be simple or complex depending on the number of scenarios considered by the system designer.

In practice, 1-GC is commonly referred to as "traffic responsive" control. Perhaps the earliest attempt at 1-GC control was a 1938 attempt at adjusting cycle length using field measurements of traffic demand [15] and analog computing equipment. The first commercially available system was the "PR System" developed in the 1950s (as reported in several papers, although few details can be found). The City of Toronto built one of the earliest area-wide 1-GC systems in the early 1960s [17]. Today, 1-GC methods have been integrated into various advanced traffic management system (ATMS) software packages. A summary of 1-GC methods is presented in Table V in the appendix.

#### 2) Second-Generation Control

Second-generation control (2-GC) measures traffic conditions and makes adjustments to COS patterns. These adjustments can be made at predefined intervals (e.g., once every five minutes) or in real time.

Some early 2-GC control methods were developed as part of the UTCS project [18], [19]. In the 1970s, SCATS was developed in Australia [20] and SCOOT in the UK [21]. SCATS and SCOOT have been widely used internationally. In the US in the early 2000s, ACS-Lite was developed to try to add 2-GC capabilities to existing traffic signal systems [22].

SCOOT predicts the amount of delay and number of stops for a typical cycle using measurements of link inflows and a platoon dispersion model [21], [23]. SCATS incorporates both traffic-responsive capabilities as well as adaptive COS adjustments [20]. Under ACS-Lite, existing COS patterns in the controller are implemented according to the time-of-day schedule, or by independent traffic responsive control [24]. ACS-Lite measures conditions for at least two cycles and incrementally adjusts the splits and offsets in the existing plan. More recent implementations have added cycle length adjustments.

Some 2-GC methods are able to optimize and implement new COS settings in real time. Several studies have explored real-time determination of splits [25]–[29]. Others have optimized each of the COS component over each intersection of the system [30]–[32]. An example is IN-TUC, which has been deployed in Greece, Germany, and the UK. The initial versions of this method optimized only splits, and later expanded to include online optimization of offsets and cycle length as well.

A summary of 2-GC control methods is presented in Table VI in the appendix.

*3) Third-Generation (Real-Time) Control*

Third-generation control (3-GC) includes methods that replace the COS paradigm with an alternative scheme for scheduling phases. Many 3-GC methods are acyclic, although some methods have been formulated for local control that is constrained within a cyclic or pseudo-cyclic system-level control. Collectively, the 3-GC methods exhibit much greater variety than the 1-GC and 2-GC methods, so they require more criteria for classification.

The first known paper on real-time traffic signal control was published by Miller [33], with another paper by Dunne and Potts [34] being published around the same time. Research under the NCHRP program explored some of these early ideas for potential real-world application [35]. The UTCS project explored 3-GC control in greater detail, under different volume conditions [13], [36]–[38], yielding three methods: SOLIS for lower traffic volumes; CYRANO for moderate traffic volume; and CIC for high traffic volume. After the end of the UTCS project, federal research in the US continued to invest in additional systems. Some of the most frequently cited of these are OPAC [39], RHODES [40], and ALLONS-D [41], [42]. Many other methods have been proposed. A summary of 3-GC control methods is presented in Table VII in the appendix.

B. *Control Scope*

ATSC methods may execute control decisions for a single intersection (local) or multiple intersections (system).

*1) Local Control*

Most early publications on real-time ATSC focused on a single intersection. Miller's seminal work [33] considered a phase-switching decision for two-phase control using the criterion of whether more delay is saved than caused by switching phases at the present moment. Dunne and Potts proposed a similar rule-based approach, also for a two-phase intersection [34]. Green and Hartley [43] considered similar rules for two, four, and eight-phase intersections.

DYPIC [44] precalculated optimal solutions for every possible link state, allowing the controller to "look up" the optimal control according to the reported conditions. DYPIC was originally intended to provide an "absolute standard of performance" for comparing other control methods.

All of the other control methods previously mentioned, including those developed under UTCS and later studies, include a method of determining the local control. In general, these methods try to predict when traffic is expected to arrive at the intersection as well as how much traffic is presently queued, and then establishes a schedule for phase switching over a planning horizon, which optimizes some performance measure. Commonly, the methods try to minimize delay or a combination of delay and stops. The methods vary in their estimation methods, selection of internal model, formulation of an optimization criterion, and algorithm for determining the schedule.

MOVA [45] is an enhancement of actuated control that uses a particular detection scheme to identify approaching and crossing traffic, developing an internal model of vehicle positions. The decision to terminate the green tries to ensure that the initial queue has cleared and then can extend the green further to minimize delay and stops.

PMSA was developed for local control under a connected vehicle environment, and uses an embedded microsimulation model to optimize delay and stops [46]. Pandit *et al.* [47] proposed an algorithm for VANETs where vehicles are aggregated into platoons, and the local level signal control problem is formulated as a job scheduling problem.

Some local control methods have features that facilitate coordination, often as an emergent property of the local control. The method of coordination is a separate criterion, discussed later in this paper.

*2) System and Hierarchical Control*

Some control methods include a framework for managing signal timing across multiple intersections. Certain control methods are scoped for individual corridors or small areas, and generally have a single layer to manage system control. Others have multiple layers that permit multiple subsystems to be managed. We term this design pattern as "hierarchical" control.

The 2-GC control methods rely on the COS framework, which is itself a manner of system control. However, different control methods differ in the complexity of the system control layer. Many 3-GC methods employ hierarchical strategies to manage local and system needs. Typically, the system component constrains the local control to facilitate coordination. For example, OPAC was originally formulated as a local control method, with certain provisions for multiple-intersection control. Coordination was later achieved through imposition of system-level constraints, using a "virtual cycle length" [48]. Similarly, the local control algorithm of RHODES (called COP) was initially developed for local intersections [49]. Additional levels in the control hierarchy were then added to manage coordination. A decision-tree method called REALBAND [50] tracks platoon movements and then requests service of specific phases to serve these platoons. PAMSCOD [51] and MMITSS [52] are extensions of these concepts employing connected vehicle data, which implement alternative system-control logic.

C. *Responsiveness*

Some ATSC systems make model-based predictions about future states. This characteristic marks another way that different systems can be classified.

*1) Reactive Control*

Reactive approaches use recent measurements of traffic. Both 1-GC and 2-GC methods tend to operate in this manner. The system state is measured over a few cycles or several minutes, and a new timing plan or set of adjustments are calculated based on the recent system state.

*2) Proactive Control*

Proactive algorithms try to determine a control decision for a predicted future traffic state, using recent measurements, sometimes in combination with historical data. Proactive methods can be divided into two subcategories based on how frequently the control decision is made: within a planning horizon (every few minutes) or as a continuous decision (each time step).

*Planning Horizon*. Many 3-GC control methods employ a planning horizon over which an optimal schedule for signal timing is determined. Some use a rolling horizon (Figure 2) wherein the control decision can be updated more frequently as the system state is refreshed.

The difficulty in measuring and predicting traffic at greater distance from the intersection tends to limit the feasible duration of the planning horizon. However, computational complexity limits the feasibility of shorter durations. Most algorithms described in the literature use longer planning horizons, with durations similar to typical cycle lengths under conventional control. Some control methods are able to vary the duration of the planning horizon. For example, ALLONS-D enlarges the existing planning horizon until all the projected queues are cleared [41]. Algorithms that use shorter horizons can tend to truncate the current phase before queues are fully cleared [53], [54]. One solution to mitigate this is to introduce a terminal cost to penalize truncation of a discharging queue.

*Continuous Decision*. Rather than determining an optimal schedule in advance, some control methods continually decide whether to extend or terminate the current phase. The decision is made at very short time resolutions, similar to actuated control (*e.g.*, 0.1–1 s). Computational efficiency is therefore critical, and such methods are often implemented as simple rule sets without complex modeling.

D. Decision Method

The next taxonomy criterion is the decision method employed as the kernel of the algorithm. Four categories were identified from the literature search: rule-based, max pressure, exact optimization, heuristic, simulation-based, and AI-based.

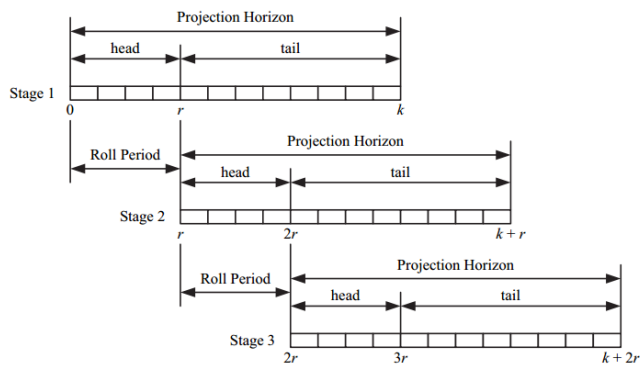

Figure 2. Planning Horizon.

*1) Rule-Based*

Rule-based methods do not employ mathematical optimization techniques, but instead use a series of relatively simple rules. These are generally limited to "if-then" rules and relatively simple calculations and value comparisons. Such methods have low computational requirements and can be executed at high frequencies but preclude more complex modeling.

*2) Max Pressure*

Max Pressure (MP) control was first proposed by Varaiya [55]. In this method, a quantity called "pressure" is defined for each phase as a function of the current traffic state and a weight. In Varaiya's initial work, queue length was selected as the weight. The group of phases having the greatest pressure is selected for service. The initial theoretical work demonstrated that MP control optimizes network throughput when the mean arrival rate is less than the service rate (i.e., undersaturated conditions).

*3) Exact Optimization Methods*

Exact optimization methods use a mathematical statement of the control problem, often incorporating a relatively complex traffic model, and derive an optimal solution. Several optimization strategies are used within these methods, which offers a means of categorizing them further.

*Dynamic Programming (DP)*. DP is a mathematical optimization technique that finds control for a dynamic system over a period to minimize cost. DP divides up the overall problem into subproblems in such a way that solving the subproblems yields global optimum performance. Unlike other recursive solutions, DP stores incremental data. If certain subproblems need to be resolved multiple times, it is better to store their parameters to avoid repeating the same calculations. Thus, DP can be seen as a sort of time-efficient recursion.

DP-based approaches for signal control use forward and backward recursion. The forward recursive algorithm usually calculates the performance function for the given parameters and stores the optimal performance value. The backward recursive algorithm backtracks the optimal decision. OPAC, COP, PAA, and SPPORT [56] use prediction algorithms to develop an arrival table, and from this information establish a signal timing plan using DP. In OPAC and COP, phase duration and sequence are optimized. PAA uses a bi-level structure to handle eight-phase control. The upper level uses a forward recursive function to determine the "barrier length" (the time when the right-of-way is transferred from phases on one street to the other street). The lower-level function returns the optimized phase sequence and phase durations.

*Mixed Integer-Linear Programming (MILP)*. Optimization of traffic signal control often requires discrete values, which has encouraged use of mixed-integer linear programming. For example, PAMSCOD [51] is a unified platoon-based formulation which performs online traffic signal optimization incorporating MILP to determine the optimal signal timing.

Computational complexity is a limitation of MILP, leading to various strategies to scale the control methods. Islam and Hajbabaie [57] limit the optimization process to the local intersection but maintain system level performance through implicit coordination. Fei *et al.* [29] extended this work to address uncertainty in traffic demand and turning rates through two-stage stochastic programming. To make optimization scalable this strategy used decomposition algorithms and coordination was maintained through cycle-offset structure. Wang *et al.* [58] formulated the MILP problem as a multi-intersection optimization problem within which vehicles can plan a path. This study uses Lagrangian decomposition with a subproblem approximation method to scale the optimization process.

*Branch-and-Bound (B&B).* This strategy renders the optimization problem into a treelike structure. The branches of the tree represent subsets of the solution set. Branches are defined by upper and lower bounds. A detailed discussion of such algorithms for two-phase control is presented by Shelby [54] Some algorithms such as ALLONS-D employ B&B in a pure form. RHODES uses B&B in a hybrid framework with DP. MILP solution methods contain B&B algorithms at their core.

*Linear Quadratic Regulator and variations.* Due to the dynamic nature of traffic, optimal feedback control methods such as model predictive control (MPC) or the linear-quadratic regulator (LQR) algorithm have been tested to solve this problem. LQR is often used for linear systems where the objective function is quadratic. INTUC used LQR and balanced the number of vehicles on links in the network by tapered split timing [25], [31]. A possible limitation of LQR is that it works on an infinite time interval, and long-term prediction of traffic is prone to error. Wang *et al.* [26] proposed the use of MPC, which works over a prediction horizon. These studies enhanced the store-and-forward mathematical model of traffic control originally developed by Gazis [59].

*4) Heuristic Methods*

Heuristic approaches employ loosely defined rules to find a solution. These solutions tend to be locally optimal rather than globally optimal. However, the need for computational efficiency, and the error inherent in predicting future traffic states can limit the practicality of identifying a global optimum solution. In other words, as the adage goes, "perfect is the enemy of good."

Several previous studies [60]–[63] used genetic algorithms (GA), which are heuristic methods that simulates the biological process of natural selection. Liang *et al.* [60] proposed a method that identifies platoons in traffic stream using connected vehicle data and determines the signal timing with a GA. GABNOR [61] uses a GA stopping criterion determined by a time limit to apply a GA solution for 2-GC. Several other heuristic based methods have also been used to optimize signal timing, such as the Box algorithm [64], Tabu search [65] and Hill-climbing [66].

*5) Simulation-based optimization*

An alternative to modeling is the use of simulation. Rather than attempting to model traffic states, potential outcomes are modeled by simulation. These may vary from rudimentary embedded simulations to the execution of simulation software. Because of the potential computational complexity, simulation-based methods are challenging to implement with real-time control, but some studies have nonetheless explored this option.

DYPIC [44] uses predetermined solutions for any given combination of demands on two phases. For every possible combination of traffic states for the simple two-phase scenario, an optimal control decision is determined by previously run simulations. The control method is therefore more similar to a lookup function.

PMSA [46] is a decentralized control method for use with V2I communication using a 15-second horizon for collecting vehicle location, speed, and headway information. Over the planning horizon, microsimulation is used to determine a solution that minimizes delay.

*6) Artificial Intelligence (AI) Based Approaches*

Recent advances in the field of AI have inspired researchers to apply it to the problem of signal control. The most common methods of AI seen in the literature on signal control are fuzzy logic, neural networks (NN), and reinforcement learning (RL).

*Fuzzy Logic.* Fuzzy logic is useful for solving nondeterministic and non-linear problems. The core concept of fuzzy logic is the replacement of the strict binary or Boolean absolute conditions of true (1) or false (0) with intermediate values representing a state that lies somewhere between the two. This is intended to emulate human decision making.

In human control of traffic signals, an experienced operator quickly reaches a decision based on their previous qualitative knowledge. Several researchers [67]–[71] have used fuzzy logic in signal control, mostly for relatively simple intersection configurations. Nakatsuyama *et al.* [72] incorporated local fuzzy control with fuzzy offset control for one-way arterial movement. Wei *et al.* [73] used fuzzy logic to control isolated intersections with through and left turn movements.

Chiu [74] proposed a self-organizing method integrating a set of 46 fuzzy decision rules to determine the signal timing. The local controller also considered the signal timing of adjacent intersections to minimize number of stops on the major street.

Fuzzy logic has been integrated with other processes in application to signal control. Xie [75] developed a RL-based framework with fuzzy logic to differentiate between congested and uncongested traffic conditions. A multi-agent independent RL method was used for arterial level signal control.

*Reinforcement Learning (RL).* RL has been used extensively in robotics and automation, and several researchers have explored its application to signal control. In RL, any decision-making element is an "agent," and the problem conditions are the "environment." The signal controller is an agent. RL-based signal control learns from the performance that results from its previous control decisions. RL-based methods can be model-based or model-free, depending on whether an agent is predicting environmental responses or using direct feedback. The control algorithms can learn both from the results of their own actions, or they can be trained by previous data. In a sense, DYPIC foreshadowed by effectively creating a training data set that was populated to include every possible traffic state.

Thorpe [76] applied RL to control of a two-phase isolated signal using a Q-table, which is computationally expensive. Abdulhai *et al.* [77] proposed a Q-learning method for system control that updated input state matrix based on projected flow from adjacent intersections, the agent also has the flexibility to update signal timing based on the projected flow.

RL methods have received considerable attention recently, and a summary of RL-based methods is presented in Table VI.

### E. Coordination Mechanism

The mechanism used for coordinating multiple intersections is a key feature of ATSC and offers another criterion for classifying ATSC methods.

#### 1) Emergent Coordination

Some ATSC methods do not include any explicit coordination method, but some have demonstrated a potential for emergent coordination that occurs as a natural consequence of the local control algorithm.

A 1966 paper [78] on fully-actuated control describes a possible "platoon carryover effect," which is perhaps the first time the notion of emergent coordination appears in the literature. However, the paper stops short of actually demonstrating this potential.

Gershenson *et al.* [79], [80] proposed an algorithm that consists of a phase-switching ruleset that tries to keep platoons together, extends green based on the balance of approaching and waiting traffic, and adjusts for downstream blockages. Another decentralized algorithm was proposed by Lämmer *et al.* [81] which used a predictive method to minimize waiting times, attempting to prevent queue formation, which also exhibits self-organizing potential.

Cesme and Furth proposed a method of self-organizing control [82] based on conventional fully-actuated control where a secondary extension is implemented to serve platoons on the major movements, subject to other criteria. Another decentralized approach proposed by Islam and Hajbabaie [57] maximizes throughput. Some other authors have described methods of "platoon accommodation" which have similar characteristics, although they focus on control of isolated intersections and do not explore system effects [83], [84]. A recent pilot study of a rule-based decentralized method [85] that scheduled the service of minor street traffic during sufficient gaps in major street traffic, which is similar in character to some of these methods.

Another implicit method of coordination is the use of upstream intersection traffic flow and signal state data to project vehicle arrivals within the rolling horizon periods of downstream intersections. OPAC-III used this form of implicit coordination, followed by several subsqequent studies [41], [57], [86]–[88]. Several RL based algorithms [75], [77], [89]–[92] used multi-agent cooperation where projected traffic flows and shared signal state data facilitated coordination.

A distinction should be drawn between *using information from* multiple intersections and *implementing control for* multiple intersections. For example, ALLONS-D is a local control method, for which a variant, ALLONS-I, utilizes additional information from adjacent intersections [41]. Within ALLONS-I, intersections can share data with each other but the control scope is local. SURTRAC [86] employs a similar approach where intersections share data with their neighbors, but the local scheduling algorithm is decentralized.

#### 2) Cycle-Offset-Split Paradigm

2-GC control uses the COS paradigm for coordination. Thus, 2-GC methods are largely focused on adjusting the governing parameters (the cycle length, offset, and splits).

SCOOT adjusts cycle length, offset, and splits according to measured demand flows and an internal platoon dispersion model. A model of cyclic traffic flows is developed that is similar to the model used in offline optimizer TRANSYT [44]. The settings are adjusted collectively to minimize a performance index consisting of delay and stops.

ACS-Lite [22] in its original form adjusts the offset and splits. The system measures arriving traffic to create cyclic profiles of the flow rate, similar to their representation in TRANSYT. The splits are adjusted to balance the level of utilization while offsets are adjusted to maximize the number of arrivals on green. Extensions of ACS-Lite, such as Kadence [93], have added an incremental cycle length adjustment.

ACDSS [27] was developed for use in midtown Manhattan, New York City, which has a famously dense and semiregular urban street grid. The system operates on two levels. In the first level, offsets and splits are selected from a library of existing timing plans. For congested conditions, offsets are selected to restrict or meter vehicle inflows. A second level adjust the splits to balance queues and reduce spillback.

#### 3) Weighted Priority

Using biasing multipliers or putting weights on phases is a time-efficient way of achieving system-wide performance in decentralized adaptive signal control.

ALLONS-D [41] used a tree-searching algorithm to minimize delay when planning phase sequence and duration. An arrival table was developed with data from upstream detectors. Weights in the objective function for certain phases are used to account for the system effects of local controller decisions. PAA [52] uses a similar approach for higher amounts of traffic, since serving the major movement under low volume is not likely to have a significant impact on performance.

#### 4) Pseudo-Cyclic Operation

Some methods eliminate the COS paradigm yet retain a control structure that functions similarly to a signal cycle (albeit more flexibly). In VFC-OPAC [48], the local cycle reference point is allowed to vary between yield points. The yield points are defined by the virtual cycle length and offsets. The local control uses a modified version of OPAC for local signal timing. A second control layer optimizes offsets, while a third control layer calculates a system-wide virtual cycle length.

#### 5) Fixed priority request

Priority control provides preferential treatment for special modes of traffic (often public transportation). Some research has proposed the handling of coordination as a priority request.

One approach is to make this a fixed (constantly recurring) request, since it is expected that such traffic will exist on the major street most of the time. Zamanipour *et al.* [94] proposed to treat coordination as a fixed priority request. PAA [52] integrated this in a bi-level optimization framework for arterial progression. Beak *et al.* [95] extended this work by adding an offset refiner.

*6) Platoon-Based Priority Request*

Another approach to handling coordination via priority request is to make the requests conditional on the existence of platoons in the traffic flow.

REALBAND [50] creates requests based on observed platoon movements. To deal with conflicts where multiple platoons are approaching an intersection on incompatible phases, a decision tree is used to determine the minimum delay option. The platoon priority requests are handled by the local control logic as an additional constraint.

Das *et al.* [96] presented the addition of platoon-based priority requests to the MMITSS system, which also handles transit and other priority requests.

*7) Iterative Approach*

Some other control methods have used an iterative process to establish coordination. Yang et al. [97] used such an approach for system control. Initially, arrival flows are estimated by a model, which is used to optimize signal timing plan. The flows are estimated again on the next step and the timing is reoptimized. The process continues until the performances converges. ALLONS-I [41] uses repeated simulation to calculate the effect of each local controller on its neighbors.

*8) Max-Plus Algorithm*

Most RL-based system-level methods use multi-agent RL. Although it is possible in theory to use a centralized agent to control all intersections, this is impractical to implement.

Van der Pol *et al.* [98] and Kuyer *et al.* [99] used an explicit coordination mechanism using the max-plus algorithm [100]. In multi-agent decision making, a coordination graph decomposes the global reward function into a sum of local terms. Each agent selects the optimal action by variable elimination. However, this variable elimination process is time consuming and not feasible for real-time control. The max-plus algorithm is an approximate alternative to variable elimination. In a series of iterations, agents share information with each other, and a final decision is reached based on the local pay-off function as well as the global payoff of the network.

## V. Additional Taxonomy Elements

This section presents other traffic signal elements that substantially affect the nature of the control, but which are secondary to defining the overall algorithm structure.

### A. Cyclic Nature

A cycle length is often used as an explicit coordination mechanism. By ensuring that all phases are served within a specific interval, it is possible to establish a reliably repeating pattern.

However, the use of a common cycle length for many intersections can be inefficient since many of the intersections may be forced to operate with a non-optimal cycle length. Another strategy is to gather intersections into smaller size groups for coordination, which may help reduce opportunities for speeding [101].

Lee *et al.* [102], [103] proposed a real-time group-based hierarchical ATSC method which uses COS. To scale the optimization, system level group-based solutions are computed in the current cycle and implemented in the next cycle. The current cycle is tuned at the local level to address real-time changes.

Most 3-GC methods are acyclic, although many incorporate other constraints having a potentially similar influence as cycle length, such as the duration of the planning horizon.

Similarities can be drawn between conventional coordination and 3-GC methods. For instance, if a local controller falls out of coordination because of a timing disruption, the controller will go into a "transition" mode where it adjusts force-off points to resynchronize. This is similar in nature to how 3-GC control schedules phase transition times in advance, although 3-GC control optimizes an explicit objective function, whereas coordination transition uses rulesets to resynchronize the signal.0

### B. Multiphase Accommodation.

Many early investigations into ATSC were limited to two-phase signals. The addition of more phases (and multiple rings) into the decision process can tremendously increase the solution space and computational complexity. The type of phase and phase sequence options is an important description of the capability of an ATSC method.

### C. Phase Sequence

ATSC methods vary in their ability to handle phase sequences (i.e., the order in which phases are served). The added ability to change the phase sequence offers another degree of freedom for the control method to adjust the signal timing to fit traffic conditions.

In practice, some agencies avoid changing phase sequences from one cycle to the next, out of caution that transgressing road user expectancy may induce violations of the signal.

### D. Detection

The type of detection, design of detection zones, and placement of detectors plays an important role in signal control, including both actuated and adaptive control.

At present, most detectors in the US are point detectors that report occupancy when a vehicle is within the detection zone. Some of these can send pulses when new vehicles are identified in the zone. Detectors may be positioned close to the stop bar, or at an upstream location. Conventional detection zone design practices are described elsewhere in greater detail [104].

Some ATSC methods are designed to use existing detection schemes to encourage implementation. ACS-Lite [99], for example, uses stop bar detection on every phase to optimize splits, and uses upstream detection to optimize offsets. The method is intended for detector locations used in practice with existing conventional control. The self-organizing method proposed by Cesme and Furth [82] uses conventional detector locations but adds a set of detectors at a location further upstream than typical practice to measure arrivals more in advance.

In future, Connected Vehicle (CV) technology is anticipated to provide continuous position and speed data. Several studies [46], [51], [52], [57], [95] have proposed methods of control using this type of data that may be able to extend the performance of local control. At present, CVs have a low market penetration rate, and it is not clear when it will become sufficiently large to enable real-time ATSC. A further challenge is the identification of non-motorized users.

Some types of detectors, such as radar and LiDAR, are able to provide vehicle position and speed data continuously [105], [106]. This effectively creates one-way V2I communication. At present, such data is rarely used directly for signal control but is instead translated into equivalent presence data for detection zones. Gates [107] and Sharma [108] proposed applications of such data for conventional signal control. Day *et al.* demonstrated the use of upstream detector data for actuated control in a pilot field study [85]. A proof-of-concept study by Shams and Day [109] showed the potential performance of using trajectory data to replace conventional stop bar detectors.

### E. Traffic Quantification (Input Data)

ATSC algorithms rely on data to measure the current traffic state and produce a decision. The type of data required by the control method offers another criterion for classification.

Some ATSC methods use flow measurements. The quintessential example is the measurement of cyclic flow profiles by SCOOT [21], [23]. Other ATSC methods develop tables of vehicle estimated times of arrival (ETAs). Both types of data are developed from detector input, with varying degrees of modeling used to project the flow pattern forward in time.

The use of disaggregated data can increase the size of the state space, significantly increasing the computational complexity. To solve this problem, several researchers have simplified arrival patterns as platoons or clusters of vehicles. The was pioneered in the offline optimizer MITROP, which used simplified flow profile shapes [110]. SURTRAC [86] and PAMSCOD [51] use clusters to lower the computational burden of the algorithm. Datesh *et al.* [111] grouped vehicles using k-means clustering for a two-phase control scheme.

### F. Multimodal Compatibility

While signal control mainly exists because of motor vehicle traffic, there are many other modes present at intersections, including pedestrians, bicycles, and transit. Emergency vehicles and freight may also be considered additional modes. ATSC methods vary in the degree to which they accommodate these and incorporate them into decision making.

Previously, researchers have investigated the operational benefits of prioritizing transit [112]–[115]. Some methods such as PAMSCOD [51] and PAA [52] allow transit priority requests as additional constraints to the optimization formulation. Other strategies [21], [25], [44], [45], [86] integrate transit movement into the algorithm structure.

In general, the handling of non-motorized traffic is an underexplored subject in ATSC. Several researchers [27], [51], [94] have explored ways to include pedestrians or bicyclists in the optimization framework, but the studies tend to focus on motorized traffic (both private vehicle and public transportation).

### G. Handling of Oversaturation

Signal timing for oversaturated networks is complex and requires extensive analysis [116], [117]. Most ATSC methods (and most conventional signal timing plans, for that matter) are designed for undersaturated conditions where intersections have adequate capacity for their demands and where it is possible to coordinate intersections for progressive traffic flow. For oversaturated networks, however, not all movements may have adequate capacity and the role of coordination is more to avoid gridlock and permit as much throughput as possible, rather than promoting smooth flow.

SCOOT [21], [23] uses specific offsets for congested conditions and also has a "gating" function that meters traffic at selected locations. IMPOST [118]–[120] controls the growth of queues on a saturated approach by metering traffic to maintain stable queues. ACDSS implements certain control strategies to manage congestion including the use of specific offsets for oversaturation. Cesme and Furth [121] proposed strategies for oversaturated signal control by preventing spillback and starvation. Temporary spillbacks were allowed at upstream intersections to prevent starvation at downstream critical intersections, and vice versa.

## VI. Trends in ATSC Research

Table II shows the year frequency distribution by publication date of selected taxonomy elements for the 84 ATSC methods selected from the literature to develop the taxonomy. The table shows the number of *control methods,* as opposed to the number of papers published. The earliest publication date was used to place each method in the table.

Early ATSC methods were mainly reactive, including 1-GC and 2-GC methods. Today, most field-deployed ATSC methods are still in this category. Considerably more research effort—at least in terms of numbers of papers in the scientific literature—has been invested in developing 3-GC control. Consequently there are more variations of 3-GC control than the earlier generations, and the number of publications on 3-GC control has greatly increased in recent years.

The implementation of 3-GC control methods was challenging in the early history of ATSC. Despite these limitations, several 3-GC methods were developed and tested. Since the year 2000, the number of decision methods tested for 3-GC has greatly increased, introducing novel concepts such as self-organizing and max pressure control. It seems that most concepts that emerge in the computer and control engineering spaces eventually find their way to a traffic signal application. AI-based approaches have recently become popular. ML-based algorithms do not require pre-specified models of the environment, and the control agents in RL can automatically learn relationships between action and reward. Modeling traffic signal control while considering traffic flow and route choice is a complicated task. It is unsurprising that methods that sidestep model specification have attracted attention. Although this is a relatively new field, there has been a lot of work in this space, and 17 out of the 84 selected ATSC methods included here involve RL, despite this being a relatively new category.

## VII. Taxonomy of Adaptive Methods

Figure 3 presents an illustration of the ATSC taxonomy that was developed by a read through the literature as described to determine the categories described earlier. Certain elements are specific to 3-GC methods (or to RL-based methods, which are a subcategory of 3-GC), and are not included. Detailed categorization of the selected methods, including these subcategory classifications, are presented by Tables III–VI in the Appendix.

Major categories in the ATSC taxonomy include the reactive or proactive nature of the control; the control generation, using the UTCS definitions; the control scope; the method of coming to a decision; and the coordination mechanism. Additional categories include the cyclic or acyclic nature of the control method; the flexibility of phase sequencing; types of input data; and type of detection required.

Traffic signal control has been evolving with the improvement in technology. Historically, developers of ATSC methods have had to balance computational and operational efficiency. Traffic responsive (1-GC) strategies used relatively simple calculations to select patterns from a preexisting library of options. Pattern-adjusting (2-GC) algorithms introduced incremental adjustments of these patterns through improved measurements and modeling techniques. Real-time control (3-GC) methods have continually introduced increasingly complex strategies for modeling and optimization.

## VIII. Conclusion

This study reviewed existing literature to develop a taxonomy of adaptive traffic signal control. This included an initial selection of over 1700 articles, of which 137 describing 86 ATSC methods were selected for a more detailed read to gather the information needed for the taxonomy. While this does not necessarily represent an exhaustive search of the literature, we believe this is a representative sample of the most widely cited and influential works in this area of study. A comprehensive review of this literature is challenging because of the variety of works reported in the literature. In fact, the tremendous amount of variety, which is continually increasing as new ideas are applied to the signal control problem, was the primary inspiration for the development of the taxonomy.

We anticipate that this taxonomy will be useful for future researchers to survey the literature on ATSC and understand the scale of the preceding work in this space. As the amount of literature continues to accumulate in this area of study, the need for such a document is increasing. In future, it is likely that additional branches and categories will be necessitated as technology improves both on the infrastructure and vehicle side. For example, in some future era, the possibilities of integrated vehicle routing [122], [123] or of automated intersection control [124], [125] may necessitate additional top level categories (i.e., additional control generations). As these developments arise, it will be useful to occasionally update and expand the taxonomy presented here.

TABLE II Distribution of Selected ATSC Algorithms by Publication Date

| Category | Generation | Decision Method | Scope | 1980 or earlier | 1981–90 | 1991–2000 | 2001–10 | 2011–22 | All Years |
|---|---|---|---|---|---|---|---|---|---|
| Reactive | 1-GC | All | S | 1 | | | 1 | | 2 |
| | 2-GC (pattern adjusting) | All | S/H | 1 | 1 | 1 | 3 | | 6 |
| Proactive | 2-GC (real-time) | All | S | | | | 3 | 9 | 12 |
| | 3-GC, Continuous Decision | Self-Organizing | L | | | | 2 | 1 | 3 |
| | | Max Pressure | L | | | | | 4 | 4 |
| | | Rule-based (others) | L | 4 | | | 2 | 2 | 6 |
| | | RL | H/L | | | 1 | 7 | 9 | 17 |
| | | Fuzzy Logic | H/L | 1 | 1 | 1 | 2 | | 5 |
| | | Total, Continuous Decision | (all) | 3 | 1 | 2 | 13 | 16 | 35 |
| | 3-GC, Planning Horizon | DP | H/L | 1 | 2 | | 4 | 3 | 10 |
| | | MILP | H/L | 1 | | | 1 | 6 | 8 |
| | | B&B | H/L | | | | 2 | | 2 |
| | | Heuristic | H/L | | | | 2 | 4 | 6 |
| | | Other Methods | H/L | | | | 3 | 2 | 6 |
| | | Total, Planning Horizon | (all) | 2 | 2 | 2 | 10 | 15 | 32 |
| | 3-GC, All Types | | (all) | 5 | 3 | 4 | 23 | 31 | 66 |
| Total, All ATSC Selected for Review | | | | 9 | 4 | 5 | 30 | 40 | 88 |

Scope: H = Hierarchical, L = Local, S = System

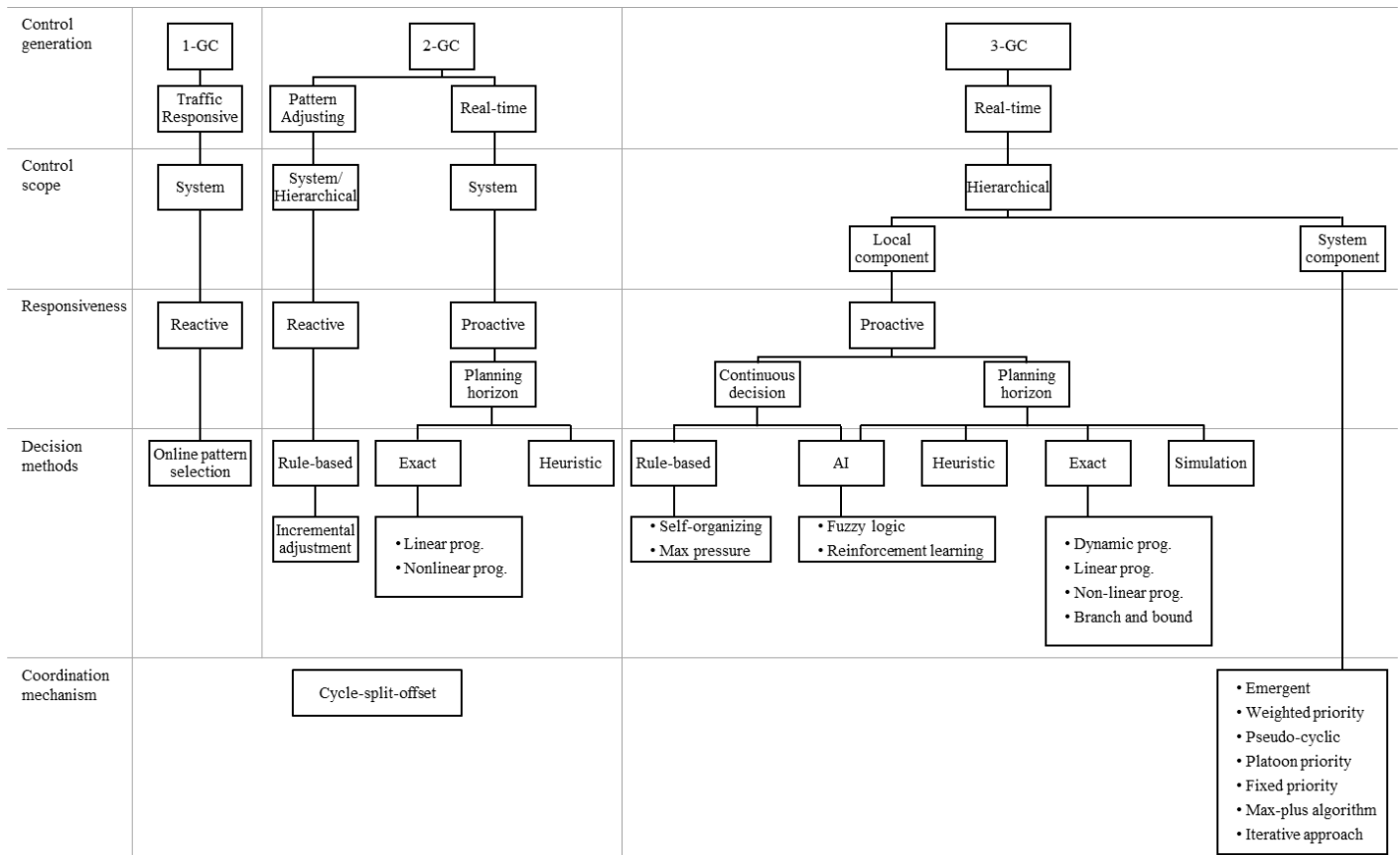

Figure 3. Visualization of the ATSC taxonomy.

X. APPENDIX

TABLE III FIRST-GENERATION ALGORITHMS

| No. | System | Scope | Traffic Quantification | Objective Function | Pattern Selection Mechanism |
|---|---|---|---|---|---|
| 1 | UTCS 1-GC [18] | S | Traffic volume | Delay | Time-of-day, operator selection, matched pattern |
| 2 | Abbas et al. [126] | S | Occupancy, delay from simulation | Delay, stops | Used genetic algorithm to select appropriate plan |

Scope: D = Distributed, H = Hierarchical, L = Local, S = System

TABLE IV SECOND-GENERATION ALGORITHMS

| No. | System | Scope | Traffic Quantification | Objective Function | Cycle Length Adjustment | Offset Adjustment | Split Adjustment |
|---|---|---|---|---|---|---|---|
| 3 | UTCS 2-GC [18] | S | Historical data | Delay | Online optimization every 5 min | CIC adjusts to accommodate queue in every cycle; if flow is over a threshold value offset will not be adjusted | CIC adjusts based on relative approach volume in every cycle |
| 4 | SCOOT [21], [23] | S | Arrival flow profile | Delay | Incrementally updated to serve critical subarea intersection | Incremental adjustment by nominal values, based on flow profile | Checks if stage change is better earlier or later |
| 5 | SCATS [20] | S | Degree of saturation, link flow | Degree of saturation | Incrementally updated based on degree of saturation | Weighted priority; offset have linear relation with cycle length | Incremental adjustment based on current degree of saturation |
| 6 | ACS-Lite [22] | S | Occupancy | Degree of saturation; arrival on green | None | Adjusts offsets with some fixed step | Incremental adjustment based on degree of saturation |
| 7 | Kadence [93] | S | Occupancy, degree of saturation | Degree of saturation, arrival on green, traffic conflict area | Incremental adjustment based on current degree of saturation | Adjust offset within a range | Incremental adjustment based on degree of saturation |
| 8 | IN-TUC [31] | S | Number of vehicles | Balance of vehicles, queue, and stops | Fixed | Fixed | Traffic control is solved as Linear Quadratic (LQ) problem, optimizes only split value |
| 9 | Wang et al. [26] | S | Detector occupancy/ Number of vehicles | Intersection utilization (ratio of throughput and capacity) | Fixed | Fixed | Traffic control is solved as Model Predictive Control (MPC) problem, optimizes only split value |
| 10 | ACDSS [27] | S | Volume, speed, occupancy, travel time | Balance queue storage ratio | Fixed | Import timing plan from library, Centrally offsets are tapered | Local controller can shift splits to prevent spillback |
| 11 | Ma et al [127] | S | Vehicle arrival flow | Delay | Fixed | Fixed | Can extend split beyond the initial split value in coordinated phases; non coordinated phases may truncate early but will not get extension. |
| 12 | McKenney et al. [28] | S | Number of vehicles | Queue length | Fixed | Fixed | Split is a function of number of vehicles and cycle length |

| No. | System | Scope | Traffic Quantification | Objective Function | Cycle Length Adjustment | Offset Adjustment | Split Adjustment |
|---|---|---|---|---|---|---|---|
| 13 | Li et al. [128] | S | Queue length | Queue spillback risk | Fixed | Fixed | Split is adjusted cycle by cycle |
| 14 | Zheng et al. [129] | S | Number of vehicles | Minimize delay | Fixed | Fixed | Formulated the problem as non-linear programming and optimized splits |
| 15 | Fei et al. [29] | S | Number of vehicles | Maximize throughput | Fixed | Fixed | This is a two-stage stochastic integer program to address uncertainty in traffic demand and vehicle turn. This method optimizes split time cycle by cycle |
| 16 | IN-TUC [31] | S | Number of vehicles | Balance of vehicles, queue, and stops | Feedback regulator optimizes cycle | Feedback regulator optimizes offset | LQR optimizes split |
| 17 | RT/ IMPOST [119] | S | Queue length | Throughput | MILP | MILP | MILP |
| 18 | Lian et al. [30] | S | Queue length | Maximize splits with weight per phase | Linear programming | Linear programming | Linear programming |
| 19 | Sun et al. [32] | S | Queue length | Terminate oversaturation period and minimize delay | Fixed | Fixed | Quasi-optimal feedback control strategy to optimize split |
| 20 | Lee et al. [102], [103] | S | Occupancy | Minimize delay | Non-linear optimization for cycle | None | Max pressure for second-by-second split decision |

Scope: H = Hierarchical, L = Local, S = System

TABLE V Third-Generation Algorithms

| No. | System | Scope | Type | Traffic Quantification | Cyclic | Objective Function | Optimization Method | Network Control |
|---|---|---|---|---|---|---|---|---|
| 21 | Miller [33] | L | CD | Arrival time | No | Delay | Rule-based | None |
| 22 | Dunne and Potts [34] | L | CD | Queue length | No | Queue length | Rule-based | None |
| 23 | UTCS 3-GC [18] | S | PH | Arrival time | No | Delay | MILP | None |
| 24 | Yu et al. [130] | L | PH | Arrival time | No | Delay | MILP | None |
| 25 | MOVA [45] | L | CD | Vehicle presence, Arrival time | No | Minimize marginal delay | Rule-based | None |
| 26 | Gershenson [80], [131] | L | CD | Number of vehicles | No | N/A | Rule-based | Emergent |
| 27 | Kwatirayo et al. [132] | L | CD | Number of vehicles | No | N/A | Rule-based | None |
| 28 | Lammer et al. [133] | L | CD | Number of vehicles | No | N/A | Rule-based | Emergent |
| 29 | Chandan et al. [134] | L | CD | Speed, position, number of vehicles | No | N/A | Rule-based | None |
| 30 | Li et al. [135] | L | PH | Queue length | No | Maximize intersection utilization | Rule-based | None |
| 31 | Bang et al. [136] | L | CD | Queue length | No | Travel time, number of stops | Rule-based | None |
| 32 | Pandit et al. [47] | L | PH | Arrival table | No | Delay | Job schedule | None |
| 33 | DYPIC [44] | L | PH | Arrival table | No | Total delay | DP | None |
| 34 | Yin et al. [137] | L | PH | Arrival table | No | Queue length | DP | None |
| 35 | Yu et al. [138] | L | PH | Number of vehicles | No | Queue length | DP | None |
| 36 | Kamal et al. [139] | L | PH | Arrival time | No | Throughput | Bi-level non-linear optimization | None |
| 37 | CRONOS [64] | L | PH | Arrival table | No | Total delay | Box algorithm | None |
| 38 | Lee et al. [62] | L | PH | Number of vehicles | No | Total delay | GA | None |
| 39 | Shenoda et al. [65] | L | PH | Arrival table | No | Stopped delay | Tabu search | None |
| 40 | PMSA [46] | L | PH | N/A | No | Delay | Simulation-based | None |
| 41 | Pappis et al. [67] | L | CD | Number of vehicles, queue | No | N/A | Fuzzy logic | None |
| 42 | Murat et al. [68] | L | CD | Longest queue on red; vehicle arrival on green; | No | N/A | Fuzzy logic | None |

| No. | System | Scope | Type | Traffic Quantification | Cyclic | Objective Function | Optimization Method | Network Control |
|---|---|---|---|---|---|---|---|---|
| 43 | Trabia et al. [70] | L | CD | Arrival time; queue | No | N/A | Fuzzy logic | None |
| 44 | Niittymäki et al. [71] | L | CD | Arrival time, pedestrian waiting time | No | N/A | Fuzzy logic | None |
| 45 | Nakatsuyama et al. [72] | L | CD | Number of vehicles, queue | No | N/A | Fuzzy logic | None |
| 46 | Talukdar et al. [106] | L | PH | Number of vehicles | No | Delay | Rule-based | None |
| 47 | OPAC [48], [140], [141] | D | PH | Arrival Table | No | Delay; or preferred function | DP | Pseudo-cyclic operation, area wide coordination |
| 48 | PRODYN [142] | D | PH | Not specified | No | Delay | DP | Emergent |
| 49 | ALLONS-D [41] | D | PH | Queue length | No | Delay | B&B | Emergent; weighted priority |
| 50 | ALLONS-I [41] | D | PH | Queue length | No | Total delay | B&B | Iterative approach |
| 51 | RHODES [40], [49], [50] | H | PH | Arrival Table | No | Total delay, queue length | DP | Platoon request |
| 52 | SchIC [143] and SURTRAC [86] | D | PH | Queue length, Arrival table | No | any preferred function can be used | DP | Emergent |
| 53 | PAA [52] | D | PH | Arrival table | No | Total delay, queue length | DP | Priority request; weighted priority |
| 54 | Katwijk [87] | D | PH | Arrival table | No | Delay; or preferred function | DP | Iterative approach |
| 55 | Yang et al. [97] | D | PH | Vehicle flow | No | Delay | DP | Iterative approach |
| 56 | PAMSCOD [51] | D | PH | Arrival table | No | Total delay, queue length | MILP | Platoon priority request |
| 57 | DC-Approach [57] | D | PH | Number of vehicles | No | Throughput with queue penalty | MILP | Emergent |
| 58 | DC-Approach [88] | D | PH | Number of vehicles | No | Throughput with queue penalty | MILP | Emergent |
| 59 | Varaiya [55] | D | PH | Queue length | No | Pressure (queue length) | Max Pressure | Emergent |
| 60 | Kouvelas et al. [144] | D | CD | Queue length | Yes | Rule-based | Max Pressure | Emergent |
| 61 | Sun et al. [145] | D | PH | upstream and downstream queue length | Yes | Rule-based | Max Pressure | Emergent |
| 62 | Levin et al.[146] | D | PH | upstream and downstream queue length | Yes | Rule-based | Max Pressure | Emergent |
| 63 | IDSTOP [14] | S | PH | Volume | Yes | Number of weighted trips | GA | COS |
| 64 | Liang et al. [60] | D | PH | Vehicle speed, position | No | Delay | Intelligent tree search, GA | Platoon-based phase sequencing and phase duration |
| 65 | Cesme et al. [82] | D | CD | Arrival Table | No | N/A | Rule-based | Emergent |
| 66 | GASCAP Uncongested flow [147] | D | CD | Number of approaching vehicles, vehicles in queue | No | N/A | Rule-based | Offset like mechanism |
| 67 | Araghi et al. [148] | D | PH | Queue length | No | Delay | Cuckoo search optimization | Emergent |
| 68 | PODE [61] | L | PH | Queue size, Arrival time | | Delay | Piecewise optimization | None |
| 69 | GABNOR [61] | S | PH | Arrival time | | Delay | GA | System level delay minimization |
| 70 | Saito et al. [66] | L | PH | Traffic condition, geometry, signal | Yes | Delay | Depth first search method and heuristic based best first-method | None |

Scope: H = Hierarchical, L = Local, S = System
Type: CD = Continuous Decision, PH = Planning Horizon
Optimization Method: B&B = Branch and Bound, DP = Dynamic Programming, GA = Genetic Algorithm, MILP = Mixed-Integer Linear Programming

TABLE VI SUMMARY OF RL-BASED ALGORITHMS

| No. | System | RL Algorithm | State | Action | Reward | Coordination |
|---|---|---|---|---|---|---|
| 71 | Thorpe et al. [76] | SARSA | Vehicle number and position | Keep or switch | Number of vehicles and number of vehicles per durations | Local controller |
| 72 | Abdulhai et al. [77] | Q-learning | Traffic volume | Keep or switch | Delay and throughput | Multi-agent cooperation |
| 73 | Xie [75] | Neural Fuzzy Actor Critic RL | queue length for each phase; signal state | Keep or switch | Throughput, queue length, residual queue | Multi-agent cooperation |
| 74 | Kuyer et al. [99] | Q function updated using DP | Vehicle position | Keep or switch | Waiting time | Integrates max-plus algorithms |
| 75 | Tantawy et al. [89] | Multi-agent RL and game theory | Queue length, phase, | Keep or switch | Delay | Multi-agent cooperation |
| 76 | Arel et al. [90] | Q-learning | Relative traffic flow | Keep or switch | Delay | Multi-agent cooperation |
| 77 | Cai et al. [149] | Approximate Dynamic Programming | Traffic arrival | Keep or switch | Queue length | Local controller |
| 78 | Balaji et al. [91] | Q-learning | Traffic flow/ queue (converted using Fuzzy) | Keep or switch | Total vehicle/ delay | Multi-agent cooperation (adjacent intersections share vehicle occupancy) |
| 79 | Khamis et al. [150] | Q function updated using DP | Queue Length | Keep or switch | Trip time, waiting time | Multi-agent cooperation |
| 80 | MASTraf [151] | Q-learning and Approximate Dynamic Programming | Number of vehicles, waiting time, signal state | Keep or switch | Throughput, waiting time | Integrated max-plus algorithm with multi-agent cooperation |
| 81 | Pol et al. [98] | Deep Q-learning network | Vehicle position; signal state | Keep or switch | Delay, decelerations, waiting time, signal condition | Integrates max-plus algorithm |
| 82 | Nishi et al. [92] | Graph convolutional neural nets | Queue length; average velocity | Keep or switch | Travel time | Multi-agent cooperation |
| 83 | Wei et al. [152] | Deep Q network | Queue length, vehicle position, signal state | Keep or switch | Queue length, waiting time, travel time, delay, queue length | Multi-agent cooperation |
| 84 | Genders [153] | nQN-TSC, RL TSC | Density, queue, signal state, signal state time | Keep or switch | delay | Local control |
| 85 | Aslani et al. [154] | Actor-critic and direct exploration | Number of vehicles on each approach | Selects green time duration from a predefined set | Total number of vehicles on all approaches | Local control |
| 86 | Guo et al. [155] | Q-function approximation | Queue length on each approach | Extend up to a predefined time or switch | Queue length | Local control |
| 87 | Tan et al. [156] | Deep RL | Average travel time and queue length per approach | Keep or switch | Queue length, delay and throughput, residual queue | Local control |
| 88 | Fei et al. [29] | Q-learning | Number of vehicles in each cell; each intersection approach has 5 cells (length 30 m) per lane | Keep or switch phase for next 3 seconds | Queue length, queue length from downstream intersection penalty for excessive delay | Multi-agent cooperation |

TABLE VII ABBREVIATIONS USED IN THIS PAPER

| Abbreviation | Definition |
|---|---|
| 1-GC | First-generation control |
| 2-GC | Second-generation control |
| 3-GC | Third-generation control |
| ACDSS | Adaptive Control with Integrated Decision Support System |
| ACS-Lite | Adaptive Control System-Lite |
| ADP | Approximate Dynamic Programming |
| ALLONS-D | Adaptive Limited Lookahead Optimization of Network Signals-Decentralized |
| ALLONS-I | Adaptive Limited Lookahead Optimization of Network Signals-Iterative |
| ANN | Artificial Neural Network |
| ATSC | Adaptive Traffic Signal Control |
| B&B | Branch and Bound |
| CIC | Critical Intersection Control |
| COP | Controlled Optimization of Phases |
| COS | Cycle Offset Split |
| CV | Connected Vehicle |
| CYRANO | CYcle-free Responsive Algorithm for Network Optimization |
| DC | Distributed-Coordinated |
| DP | Dynamic Programming |
| DYPIC | Dynamic Programmed Intersection Control |
| GA | Genetic Algorithm |
| GABNOR | Genetic Algorithm Based Network Optimization in Real-time |
| GASCAP | Generalized Adaptive Signal Control Algorithm Project |
| I2V | Infrastructure-to-Vehicle |
| IMPOST | Internal Metering Policy to Optimize Signal Timing |
| INTUC | Integrated Traffic-responsive Urban Control |
| LP | Linear Programming |
| LQR | Linear Quadratic Regulator |
| MASTraf | Multi-Agent System for network wide Traffic signal control with dynamic coordination |
| MDP | Markov Decision Process |
| MILP | Mixed Integer Linear Programming |
| MOVA | Microprocessor Optimized Vehicle Actuation |
| MPC | Model Predictive Control |
| NCHRP | National Cooperative Highway Research Program |
| NLP | Non-Linear Programming |
| NN | Neural Network |
| OPAC | Optimization Policies for Adaptive Control |
| PAA | Phase Allocation Algorithm |
| PAMSCOD | Platoon-based Arterial Multi-modal Signal Control with Online Data |
| PI | Performance Index |
| PMSA | Predictive Microscopic Simulation Algorithm |
| PROTRACTS | Purdue Real-Time Offset Transitioning Algorithm for Coordinated Traffic Signals |
| RHODES | Real-Time, Hierarchical, Optimized, Distributed and Effective System |
| RL | Reinforcement Learning |
| SCAT | Sydney Cooperative Adaptive Traffic |
| SCOOT | Split, Cycle, and Offset Optimization Technique |
| SOLIS | Signal Optimization using LInk Signatures |
| SPPORT | Signal Priority Procedure for Optimization in Real-time |
| SURTRAC | Scalable Urban Traffic Control |
| TRANSYT | TRAffic Network StudY Tool |
| UTCS | Urban Traffic Control System |
| V2I | Vehicle-to-Infrastructure |
| V2V | Vehicle-to-Vehicle |
| VFC-OPAC | Virtual Fixed Cycle OPAC |